\documentstyle [aps,twocolumn]{revtex}

\begin{document}
\title{Multifractality:
generic property of eigenstates of 2D disordered metals}

\author{Vladimir I. Fal'ko $^{1,3}$ and K.B. Efetov$^{2,4}$}

\address{$^1$ Max-Planck-Institut f\"ur Fesk\"orperforschung
Heisenbergstr. 1, 70569 Stuttgart, Germany \\\
$^2$ Max-Planck-Institut f\"ur Physik Komplexe
Systeme, Heisenbergstr. 1, 70569 Stuttgart, Germany \\\
$^3$ Institute of
Solid State Physics RAS, Chernogolovka, 142432 Russia\\\
$^4$ Landau Institute for Theoretical Physics, Moscow, Russia}
\date{\today{}}
\maketitle

\begin{abstract}

The distribution function of local amplitudes of eigenstates of a
two-dimensional disordered metal is calculated.  Although the
distribution of comparatively small amplitudes is governed by laws
similar to those known from the random matrix theory, its decay at
larger amplitudes is non-universal and much slower.   This leads to
the multifractal behavior of inverse participation numbers at any
disorder.  From the formal point of view, the multifractality
originates from non-trivial saddle-point solutions of supersymmetric
$\sigma$-model used in calculations.

\end{abstract}
\pacs{73.20.Dx,71.25.-s,05.45.b}

Fractality of some unusual objects of condensed matter physics has
been intensively discussed during the last decade \cite{Mandel}.  This
notion has also penetrated to the theory of disordered systems for
characterizing the vicinity of the localization transitions.  For
example, using the coefficients of the inverse participation numbers
(IPN) known from the renormalization group treatment in $2+\epsilon$
dimensions \cite{Wegner} Castellani and Peliti \cite{Cast} have
suggested the multifractal structure of wave functions at the
localization threshold.  Later, due to considerable efforts in numerical
simulations, it has been recognized that, at criticality, the wave
functions are really multifractal which gives rise to a spectrum of
critical exponents \cite{Jan}.

The concept of multifractality of wave functions has been introduced
\cite{Cast,Jan} as a way to characterize their complexity in the
pre-localized regime and originates from the numerically observed
non-trivial dependencies of the inverse participation numbers $t_n
(V)$ on
the volume $V$ of a system.  By definition, the latter are the moments
of the distribution function $f(t)$ of local amplitudes $t\equiv |
\psi ({\bf r})|^2$ of wave
functions at an arbitrary point ${\bf r}$ inside a sample,
\begin{equation}
\label{e1}
$$t_n=\int_0^{\infty}t^nf(t)dt,\;$$
\end{equation}
$$f(t)=\left(V\nu\right)^{-1}\left\langle \sum_{\alpha}\delta (t-|
\psi_{\alpha}({\bf r})|^2)\delta (\epsilon -\epsilon_{\alpha})\right
\rangle $$
where $\psi_{\alpha}\left({\bf r}\right)$ and $\epsilon_{\alpha}$ are
eigenfunctions and eigenvalues of state $
\alpha$ of
a confined system.  The symbol $\left\langle \right\rangle $ stands for
averaging over
disorder; $\nu$ is the average density of states.  The distribution
function and the wave functions are properly normalized such that
$t_0=1$ and $t_1\equiv V^{-1}.$

The inverse participation numbers indicate very sensitively the
degree of disorder-induced-localization of states through their
dependence $t_n(V)$ on the volume of the system.  In a pure metal or a
ballistic chaotic box where the wave functions extend over the whole
system one has

\begin{equation}
\label{e2}
$$t_n\propto 1/V^n.$$
\end{equation}
If disorder makes the localization length $L_c$ much shorter than the
sample size $L\sim V^{1/d}$, the coefficients $t_n$ are insensitive to $
L$.
However, a very interesting information about the development of
localization comes from the analysis of $t_n(V)$ for small samples with
$L<L_c$.

The latter situation can occur in the vicinity of a
localization-delocalization transition, when the length $L_c(\epsilon
)$ can be
larger than $L$.  In this critical regime, the multifractality is a
manifestation of pre-localization in a piece of a matter which still
dominantly shows the metallic properties.  In the language of the
coefficients $t_n$, this is described as

\begin{equation}
\label{e3}
$$Vt_n\left(\epsilon\right)\propto L^{-\tau\left(n\right)},\qquad\tau\left
(n\right)=(n-1)d^{*}\left(n\right),$$
\end{equation}
where $d^{*}\left(n\right)$ differs from the physical dimension $d$ of the
system
and is a function of $n.$ This function gives the values of the fractal
dimensions $d^{*}\left(n\right)$ for each $n.$ For ballistic waves in a
billiard, one
would have $d^{*}\left(n\right)\equiv d.$ Once we assume that the envelope
of a
typical wave function at the length scale shorter than $L_c$ obeys a
power law $\psi (r)\propto r^{-\mu}$ with a single fixed exponent $
\mu <d/2$
\cite{Schreiber,Cast}, the set of IPN's shows $d^{*}=d-2\mu$ different
from $d$ but the same for all $n>d/\left(2\mu\right)$.  This is when one
speaks
\cite{Schreiber} of a fractal behavior with the fractal dimension $
d^{*}$.

If $d^{*}\left(n\right)$ is not a constant, this signals about some more
sophisticated structure of wave functions.  They can be imagined as
splashes of multiply interfering waves at different scales and with
various amplitudes, and possibly, with a self-similarity characterized
by a relation between the amplitude of a local splash of a wave
function, $t$, and the exponent $\mu (t)$ of the envelope of its extended
power-law tail.  The spatial structure of wave functions in the
vicinity of the localization transition was a subject of extensive
numerical studies \cite{Aoki,Jan,Schreiber}.  To distinguish between
the fractal and multifractal behavior in numerics, one has to detect
and compare the values $d^{*}(n)$ coming from, at least, several lowest
IPN's.  In the pioneering works \cite{Aoki}, the assertion about the
fractality at criticality has been made on the basis of studies of
$d^{*}\left(2\right)$, and a non-trivial dependence of dimensionality $
d^{*}\left(2\right)$ on
disorder in 2D metals \cite{Schreiber} has been observed.  Later on,
numerical study of single-electron states at the transition between
the plateaus in the regime of the Quantum Hall Effect (QHE) \cite{Jan}
has established different fractal $d^{*}(n)$ for different IPN's, which
supported the idea about the multifractality in the critical regime.

The goal of this Letter is to demonstrate that the multifractality of
wave functions is a {\it generic property of 2D disordered systems\/} as
soon as the sample size $L$ does not exceed the formal localization
length $L_c$.  In some sense, the two-dimensional system is critical not
only in the QHE regime, and we consider the limit of relatively weak
magnetic fields.  If disorder is weak (the diffusion coefficient $
D$ is
large, so that $2\pi\nu D\gg 1$), the localization length $L_c$ is
exponentially large
and, therefore, the system is an ideal object to observe the
multifractality at a large scale.  To anticipate a little, our
calculation gives

\begin{equation}
\label{e4}
$$d^{*}(n)=2-n\left(4\pi^2\nu D\right)^{-1}$$
\end{equation}
which can be associated with a power-law tail of the envelope with
the exponent $\mu (t)<1$.  The result of Eq.  (\ref{e4}) is essentially
non-perturbative and cannot be obtained by any expansion in diffusion
modes usual for describing weak localization effects.

To get this result,  we use below the supersymmetry technique
\cite{Efetov}.  Analytical study
of wave functions with this method has started recently \cite{EP}.
It was discovered that the distribution of local amplitudes of waves
in a very small - zero-dimensional (0D) - disordered conductor or a
chaotic cavity obeyed the same laws as those predicted in the
random matrix theories \cite{Brody}.  One of the achievements of the
method is that it makes possible to calculate the distribution function
of local amplitudes $f(t)$ as a whole \cite{FalkoEf}, rather than to
reconstruct \cite{EP} it from the full set of IPN's.  This proves to
be very helpful for calculations beyond 0D.

Using the supersymmetry technique, one reduces the calculation of
the distribution function $f\left(t\right)$ to evaluation of a functional
integral
with the free energy functional

\begin{equation}
\label{e5}
$$F[Q]={{\pi\nu}\over 8}\int {\rm S}{\rm t}{\rm r}\left[D(\nabla Q\left
({\bf r}\right))^2-2\gamma\Lambda Q\left({\bf r}\right)\right]d{\bf r}
,$$
\end{equation}
where $D$ is the diffusion coefficient, $\gamma$ is a level width,.
The notations for the supermatrices $Q$ and $\Lambda$, and for the
supertrace
'Str' are the same as in Ref.
\cite{Efetov}.  A general form of the functional integral valid
for an arbitrary magnetic field can be found in Ref.  \cite{FalkoEf}.
In what follows, we calculate the distribution function $f\left(t\right
)$
for the unitary ensemble, although the results for the orthogonal and
symplectic
ensembles are similar \cite{Tobepublished}.  The unitary ensemble is
technically most simple.  At the same time, it can be a quite
interesting one, due to its relation \cite{ArMir} to the recently
suggested description of some strongly correlated systems, such as
high $T_c$ materials or $2D$ electron gases at even-denominator filling
factors, in terms of fluctuating guage fields  \cite{Anderson,Halperin}.

For the unitary ensemble, the function $f\left(t\right)$ can be
equivalently
rewritten in a somewhat simpler form

$$f(t)=\lim_{\gamma\to 0}\int {\rm D}Q\int{{d{\bf r}}\over {4V}}{\rm S}
{\rm t}{\rm r}(\pi_b^{(1)}Q({\bf r}))$$

\begin{equation}
\label{e6}
$$\times\delta\left(t-{{\pi\nu\gamma}\over 2}{\rm S}{\rm t}{\rm r}
(\pi^{(2)}_bQ({\bf r}_o)\right)\exp\left(-F\left[Q\right]\right)$$
\end{equation}
where the $\pi_b^{(1,2)}$ select from $Q$ its boson-boson sector,
$$\pi^{(1)}_b=\left(\matrix{\pi_b&0\cr
0&0\cr}
\right),\;\pi^{(2)}_b=\left(\matrix{0&0\cr
0&\pi_b\cr}
\right),\;\pi_b=\left(\matrix{0&0\cr
0&\tau_0\cr}
\right),$$
and $\tau_{0,3}$ are the Pauli matrices.  The limit $\gamma\to 0$ in Eq.
(\ref{e6})
corresponds to a closed system with elastic scattering only, and one
could, first, calculate the integral over $Q({\bf r})$ and take the limit $
\gamma\rightarrow 0$
at the end.  However, it is better to get rid of the parameter $\gamma$ at
an earlier stage by integrating, first, over the zero space harmonics
of $Q$.

If we restrict ourselves with the integration over this harmonics
only, we arrive at the $0D$ result of Ref.  \cite{FalkoEf}.
Nevertheless, this would not be enough for probing the states which
are the precursors of localization, since the latter is related to the
non-homogeneous fluctuations $Q({\bf r})$ \cite{Efetov}.  Moreover, as has
been recently shown by Muzykantskii and Khmelnitskii \cite{Dima},
the treatement of rare localization events in the metallic regime
(such as long-living states in an open mesoscopic conductor) can be
advanced using the saddle-point solutions of the supersymmetric field
theory.  Thus, we also account for spatial variations of $Q$
non-perturbatively, using the saddle-point method like in Ref.
\cite{Dima}, so that the integration over the zero harmonics is only
our first step.

To realize this plan, we represent the matrix $Q$ as
$$Q\left({\bf r}\right)=V\left({\bf r}\right)\Lambda\bar {V}\left(
{\bf r}\right),\quad V\left({\bf r}\right)\bar {V}\left({\bf r}\right
)=1,$$
and substitute $V\left({\bf r}\right)\rightarrow V\left({\bf r}_o\right
)V\left({\bf r}\right)$ and $Q({\bf r})\to V\left({\bf r}_o\right)
Q\left({\bf r}_o\right)\bar {V}\left({\bf r}_o\right)$.  In
terms of new variables $%
V\left({\bf r}\right)$ and $Q\left({\bf r}\right)$, the first (gradient)
term in
Eq.  (\ref{e5}) preserves its form, but now the condition $Q\left(
{\bf r}_o\right)=\Lambda$
has to be fulfilled.  The second term in Eq.  (\ref{e5}) transforms
into
$$F_2=-{{\pi\nu}\over 4}\int d{\bf r}{\rm S}{\rm t}{\rm r}(\tilde {
Q}_oQ\left({\bf r}\right)),\quad\tilde {Q}_o=\bar {V}\left({\bf r}_
o\right)\Lambda V\left({\bf r}_o\right).$$
A corresponding substitution can be done in the pre-exponential of Eq.
(\ref{e6}), too.  This enables us to integrate over $Q_o.$ The limit $
\gamma\rightarrow 0$
drastically simplifies the computation, since it makes the essential
values of the variable $\theta_{1o}$ parametrizing the non-compact sector
of
the supermatrix $Q_o$ as large as $\cosh \theta_{1o}\sim 1/\gamma
.$ Performing a standard
integration over the elements of the supermatrix $Q_o$ - the procedure
being a slight modification of the calculations of Ref.  \cite{FalkoEf}
- we reduce Eqs.  (\ref{e5}, \ref{e6}) to the form

\begin{equation} \label{e10}
$$f(t)={1\over V}{{d^2}\over {dt^2}}\left\{\int_{Q\left({\bf r}_o\right
)=\Lambda}\exp\left(-\tilde F\left[t,Q\right]\right){\rm D}Q({\bf r}
)\right\}$$
\end{equation}
where the free energy functional $\tilde {F}\left[t,Q\right]$ is determined
by
\begin{equation}
\label{e11}
$$\tilde {F}[t,Q]=\int d{\bf r}{\rm S}{\rm t}{\rm r}\left({{\pi\nu
D}\over 8}(\nabla Q\left({\bf r}\right))^2-{t\over 4}\Lambda\Pi Q(
{\bf r})\right).$$
\end{equation}
The matrix $\Pi$ selects from $Q$ its non-compact boson-boson sector
$Q_b=V_b\Lambda\bar {V}_b$:
\begin{equation}
\label{e12}
$$\Pi =\left(\matrix{\pi_b&\pi_b\cr
\pi_b&\pi_b\cr}
\right),\;V_b=\exp\left(\matrix{0&u\pi_b\theta_1/2\cr
u^{+}\pi_b\theta_1/2&0\cr}
\right)$$
\end{equation}
and $u=\exp (i\chi\tau_3)$, where $\theta_1\ge 0$ and $0\le\chi <2
\pi$.  The 0D result can be
obtained just by putting $Q\left({\bf r}\right)=\Lambda$ for all $
{\bf r}$, and one ends with the
Porter-Thomas distribution $f^{(0)}(t)=V\exp (-Vt)$ for the unitary
ensemble which gives the IPN which obey Eq.  (\ref{e2}) and show
neither multifractal nor fractal behavior.

Nonetheless, this would be only an approximate procedure, because
the value $Q\left({\bf r}\right)\equiv\Lambda$ does not correspond to the
minimum of the
functional $\tilde {F}\left[t,Q\right]$ when $t\ne 0$.  The second term
acts on the matrix $
Q$
as if an external field tended to ''align'' the matrix $Q_b$ along such a
direction that $\theta_1\rightarrow\infty .$ However, the condition $
Q\left({\bf r}_o\right)=\Lambda$ prevents from
that, and, as in Ref.  \cite{Dima}, the minimum corresponds to a
non-homogeneous configuration of a finite $Q_b\left({\bf r}\right)$.

The solution $\theta_t$ corresponding to the minimum of the functional
$F\left[t,Q\right]$ can be found after substituting Eqs.  (\ref{e12}) into
(\ref{e11})
and varying $\theta_1$ under the condition ${\bf n}\nabla\theta_t=
0$ at the boundary of a
sample, and $\theta_t\left({\bf r}_o\right)=0$ in the origin.  As a result,
we obtain

\begin{equation}
\label{e13}
$$\Delta\theta_t({\bf r})=-{t\over {\pi\nu D}}\exp\left(-\theta_t\left
({\bf r}\right)\right),\;\chi\left({\bf r}\right)=\pi ,$$
\end{equation}
where $\Delta$ is the Laplacian.  This saddle-point equation is somewhat
different to that discussed in Ref.  \cite{Dima}.  The solution of Eq.
(\ref{e13}) has to be substituted into the free energy functional
$\tilde {F}\left[t,Q\right]$ which takes the form

\begin{equation}
\label{e14}
$$F_t=\int\left({{\pi\nu D}\over 2}(\nabla\theta_t)^2+t\exp\left(-
\theta_t\right)\right)d{\bf r}$$
\end{equation}
and gives the final result with an exponential accuracy.

In principle, it can be important to take into account fluctuations in
the vicinity of the extremum.  In the lowest order, these
fluctuations contribute to the pre-exponential.  As concerns higher
order terms, one should estimate them to be sure about the
corrections to the approximations we made.  This can be done in an
invariant form representing $Q\left({\bf r}\right)$ as
\begin{equation}\label{e15}
$$Q({\bf r})=V_t({\bf r})\Lambda{{\left(1+iP\left({\bf r}\right)\right
)}\over {\left(1-iP\left({\bf r}\right)\right)}}\bar {V}_t({\bf r}
),\;P\equiv\left(\matrix{0&B\cr
\bar {B}&0\cr}
\right)$$
\end{equation}
The supermatrix $V_t\left({\bf r}\right)$ is determined by Eq. (\ref{e12})
with the
optimal solution $\theta_t$ substituted as the variable $\theta_1$, and $
P$ is treated as
a fluctuation around the inhomogeneous optimal solution.  To evaluate
the contribution of fluctuations $P$, one has to substitute $Q\left
({\bf r}\right)$ from
Eq.  (\ref{e15}) into Eqs.  (\ref{e10}, \ref{e11}) and, then, expand them
in $P.$ In the zeroth order in $P$, Eq.  (\ref{e15}) corresponds to the
optimal solution which defines $F_t$, Eq.  (\ref{e14}).  After expanding
the action $\tilde {F}[t,Q]$ in series on the perturbation $P$, one can
check that
the above choice of $\bar {V}_t$ cancels the linear terms $F^{\left
(1\right)}$, and
$\tilde {F}[t,Q]=F_t+F^{\left(2\right)}+F^{\left(3\right)}+F^{\left
(4\right)}+...$.  At this stage of calculating the
integral in Eq.  (\ref{e10}), one has to keep both $F_t$ and $F^{\left
(2\right)}$ in the
exponent $e^{-\tilde {F}[t,Q]}$, but expand it with
respect to other higher order
terms on $P$.  Then each term of this expansion is given by a
Gaussian integral and can be evaluated using Wick's theorem.

Estimating the contribution of the higher-order terms, one can see
that they are not important as long as the sample size $L\sim V^{1
/d}$ is
smaller than the localization length $L_c$ which holds in 1D and 2D
samples.  As concerns the 3D case, the fluctuations are always small
when disorder is weak.  In all the cases, the contribution from $F_
t$
dominates, but we are also aware of what comes from square
terms $F^{\left(2\right)}$.  Calculating Gaussian integrals one
obtains for the pre-exponential $J$

\begin{equation}\label{expJ}
$$J=\exp\left\{{1\over 2}\sum_n\ln\left(\chi_f^4(n)/\prod_{\alpha
,\beta}\chi_b(\alpha ,\beta ,n)\right)\right\}.$$
\end{equation}
$J$ differs from unity, since the optimal fluctuation breaks the
symmetry between anti- (f) and commuting (b) components of $P$
and splits the spectra $\{\chi_f(n)\}$ and $\{\chi_b(\alpha ,\beta
,n)\}$ of their eigen-modes.
The set of four-fold degenerate modes $\chi_f$ and non-degenerate $
\chi_b$ are
defined from the 'Schroedinger equations'
$[-2\pi\nu D\Delta +U-\chi ]s=0$, where the effective potentials
$U_{(f,b)\alpha}=2\pi\nu Dk_{(f,b)\alpha}(\nabla\theta_t)^2+t\kappa_{
(f,b),\alpha}\exp (-\theta_t)$ ($k$ and $\kappa$ are real
numbers \cite{Tobepublished}) obey the sum rule $\sum U_{b,\alpha}
=4U_{f,\alpha}$.

Until now all the manipulations did not depend on the dimensionality.
However, solving Eq.  (\ref{e13}) has to be done for each
dimensionality separately.  Results for 1D and 3D will be presented
elsewhere \cite{Tobepublished}.  Below, we discuss the 2D case.  The
use of the $\sigma$-model limits the length-scales from below by the value
of the mean free path $l$.  Therefore, when solving Eq.  (\ref{e13}), we
cut off the radii $r<l$ and replace the conditions $Q=\Lambda$ ($\theta_
t=0$) at the
origin by the same requirement at $r=r_0\sim l$, provided the dependence
of the finally derived optimal action $F_t$ on the cut-off length is
weak.  For the sake of simplicity, we consider the sample in the
form of a disk of a radius $L$, $\ln (L/l)\ll (2\pi\nu D)^2$, and place the
observation point ${\bf r}_o$ in its center.  This geometry allows us to
seek
for an axially symmetric solution of Eq.  (\ref{e13}) which, with the
parameter $\rho =\sqrt {2\pi\nu D/tl^2}$, takes the exact form
$$e^{-\theta_t}=\left[{{2(l/r)^{1-A}[\sqrt {({1\over {A\rho}})^2+1}
+1]}\over {~[\sqrt {({1\over {A\rho}})^2+1}+1]^2-({1\over {A\rho}}
)^2({r\over l})^{2A}}}\right]^2\sim\left({l\over r}\right)^{2\mu}$$
where $A=1-\mu$ is determined from the boundary conditions as

\begin{equation}\label{Ad2}
$$\sqrt {A^2+\rho^{-2}}+A={{(L/l)^A}\over {\rho}}\sqrt {{{1+A}\over {~
1-A}}}.$$
\end{equation}
Eq.  (\ref{Ad2}) has positive roots
if $\rho >\ln{L\over l}\gg 1$, which limits the
amplitude $t$ by the value $(\lambda_Fl)^{-1}$, the density of a plane wave
forward-and-backward scattered within the mean free path length.  In
this limit $\mu\ll 1$, which is the reason for the power-law asymptotics
of the optimal $\exp (-\theta_t)$.  The same conditions also simplify the
search
of roots of Eq.  (\ref{Ad2}) and reduce it to the equation on $\mu$:
\begin{equation}\label{T}
$$\mu ={{z(T)}\over {2\ln (L/l)}};\;ze^z=T\equiv{{tV\ln (L/l)}\over {
2\pi^2\nu D}}.$$
\end{equation}

The leaading terms of the optimal free energy $F_t$ can be found as
$$F_t\approx (2\pi )^2\nu D\{\mu +\mu^2\ln{L\over l}\},$$
and the logarithmic dependence of $F_t$ on the ratio $L/l$ in all the cases
justifies the cut-off procedure and makes meaningful the use of the
derived formulae for an arbitrary position of the observation point
and the shape of the sample.

One can also evaluate the pre-factor $J$ from Eq.  (\ref{expJ}) and
find that $J=1+0(T^2$ ) at $T\ll 1$, and $J\propto\mu\exp (\mu\ln{
L\over l})\sim T\ln T$ when
$T\gg 1$.  This confirms our previous statement that the optimal action
dominates over other contributions and enables us to find the form of
the distribution function $f^{(2)}(t)$ for small, $t\equiv |\psi |^
2<$, and for large,
$t>2\pi^2\nu D/(V\ln{L\over l})$, amplitudes with the exponential accuracy:

\begin{equation}\label{f2d}
$$f^{(2)}\approx V\left\{\matrix{\exp\left(-Vt[1-T/2]\right),\quad
T\ll 1\cr
\exp\left(-{{\pi^2\nu D}\over {\ln (L/l)}}\ln^2T\right),\quad T\gg
1.\cr}
\right.$$
\end{equation}
{}From this expression, one can conclude that disorder makes the
appearence of high-amplitude splashes of wave functions much more
probable than one would expect from the 0D Porter-Thomas formula.
Obviously, this is a localization effect.  But the tails of the states
do not decay exponentially, as one would expect
for a particle localized in a
quantum well or in a 1D wire beyond the localization length:  Even in
the asymptotic regime, the
size $L$ of the system influences the distribution.
The splashes look as if they were formed by focusing the waves by
those
rare configurations of scatterers which play the role of 'parabolic'
mirrors.  The structure of these states can be anticipated from the
way how their distribution function feel the boundary.  Since
this comes in through the form of the optimal fluctuation, we believe
that they have the power-decaying tails,
$|\psi ({\bf r})|^2\sim e^{-\theta_t(r)}\approx (l/r)^{2\mu}$
($\mu <1$), which approach the limiting $r^{-2}$ dependence for the highest
amplitudes $t\sim (l\lambda_F)^{-1}$.

Moreover, the form of IPN's, $t_n$ derived on the basis of Eq.
(\ref{f2d}) shows a scaling with the size of a system which forces us
to assign them a multifractal nature.  To find the moments $t_n$
accurately enough, we have to take into account that, despite of that
the cross-over to the 0d case looks like a formal limit $T(t)\to 0$, the
Porter-Thomas statistics fails unless the condition $tV\ll\sqrt {2
\pi\nu D}$ is
satisfied.  Hence, only the first few ratios $t_n$, $2\le n\ll\sqrt {
2\pi\nu D}$, can be
estimated by using a finite polynomial expansion of $f(t)$ into the
series on $T$.  The result would look as a correction to the values
expected for chaotic quantum billiards \cite{Mirlin}.  Alternatively,
we derive the higher order IPN's from the intermediate result of Eq.
(\ref{e10}),
$$t_n={{n(n-1)}\over {V^n}}\int dtJ(t)\exp\left\{-F_t+(n-2)\ln (tV
)\right\},$$
by the saddle-point method.  The moments $t_n$ calculated in both ways
are in a good agreement with each other and, in the leading order,
take the form
$$t_n\approx{{\min \{n!,[2\pi\nu D/\ln{L\over l}]^n\}}\over
{l^{2\delta}}}\left
({1\over V}\right)^{n-\delta},\quad\delta\approx{{n(n-1)}\over {~8
\pi^2\nu D}}.$$

To summarize the result of the derived above exact statistics of
amplitudes of wave functions in mesoscopic disordered conductors, we
end with such a volume-dependence of the inverse participation
numbers $t_n$ which manifests the multifractal behavior of quantum
states, Eq.  (\ref{e3}).  Multifractality seems to be the generic
property of 2D disordered systems.  In the unitary ensemble, the
multifractal dimensions are $d^{*}=2-n(4\pi^2\nu D)^{-1}$, and all over the
metallic regime, the dependence of $d^{*}$ on $n$ and the level of
disorder,
Eq.  (\ref{e4}), is exact and can be a subject of numerical
verifications.

We thank D. Khmelnitskii for sending us the preprint \cite{Dima}
prior to publication.


\begin{references}

\bibitem{Mandel} B.B.  Mandelbrot:  The fractal Geometry of Nature,
W.H.  Freeman, San Francisco (1983); R.  Benzi {\it et al\/}, J.  Phys.
A{\bf 17},
3521 (1984); J.  Phys.  A {\bf 18}, 2157 (1985); T.C.  Halsey, {\it et
al\/} Phys.
Rev.A {\bf 33,} (1986)

\bibitem{Wegner}  F. Wegner, Z. Phys. B{\bf 36}, 209 (1980)

\bibitem{Cast}  C. Castellani, L. Peliti, J. Phys. A {\bf 19,} L429 (1986)

\bibitem{Jan} M.  Janssen, O.  Viehweger, U.  Fastenrath and J.
Hajdu, Introduction to the Theory of the Integer Quantum Hall Effect,
VCH, Weinheim (1994), and Refs. therein

\bibitem{Aoki}  H.Aoki, J.Phys.C{\bf 16,} L205 (1983); Phys. Rev.B{\bf 33,}
7310 (1985); B. Kramer {\it et al\/}, Surf. Sci. {\bf 196,} 127 (1988)

\bibitem{Schreiber}  M. Schreiber, J. Phys. C {\bf 18,} 2493, (1985)

\bibitem{Efetov}  K.B. Efetov, Adv. Phys. {\bf 32}, 53 (1983)

\bibitem{EP}  K.B. Efetov, V.N.Prigodin, Phys. Rev. Lett. {\bf 70}, 1315
(1993)

\bibitem{Brody}  T.A. Brody {\it et al\/}, Rev. Mod. Phys. {\bf 53}, 385
(1981)

\bibitem{FalkoEf}  V.I. Fal'ko, K.B. Efetov, Phys. Rev. B {\bf 50}, 11267
(1994)

\bibitem{Tobepublished}  V.I. Fal'ko and K.B. Efetov, unpublished

\bibitem{ArMir} A.G.  Aronov, A.D.  Mirlin, Phys.  Rev. B {\bf 49}, 16609
(1994)

\bibitem{Anderson} P.W. Anderson, Science {\bf 235}, 1196 (1987)

\bibitem{Halperin} B.I. Halperin, P.A. Lee and N. Read, Phys. Rev. B
{\bf 47}, 7312 (1993)

\bibitem{Dima}  B.A. Muzykantskii and D.E. Khmelnitskii, preprint

\bibitem{Mirlin}  Y.V. Fyodorov and A.D. Mirlin, unpublished

\end{references}
\end{document}